\documentstyle[11pt,paspconf,epsf]{article}
\begin{document}
\title{Keck Spectropolarimetry of BALQSO}
\author{P. M. Ogle}
\affil{105-24 California Institute of Technology, Pasadena, CA 91125}

\begin{abstract}

Spectropolarimetry of broad absorption line quasi-stellar objects with the
W. M. Keck Telescope is  used to put constraints on their nuclear structure. 
If the continuum polarization is due to electron (or dust) scattering, then 
the scattered and direct rays probe the nucleus along multiple lines of sight,
which suffer differing amounts of absorption. Multiple sources of polarized 
light contribute to the quasar emission, and the relative sizes and symmetries
of these sources can be ascertained. The size of the electron scattering 
region relative to continuum source, broad emission line region, and broad 
absorption line region is determined using spectropolarimetry. The geometries 
of the  scattering and broad absorption line regions are discussed in an 
attempt to discriminate between polar and equatorial outflows. The variation
of polarization with velocity in the broad absorption troughs may also be used
to study the dynamics of the absorbing gas.

\end{abstract}

\section{Observations}
Observations were made with the Low-Resolution Imaging Spectrograph on the
W. M. Keck 10m Telescopes of 19 broad absorption line quasi-stellar objects
(BALQSO) with a variety of absorption trough morphologies, with $m_V \sim 18$,
and redshift $z=1.6-3.5$. The spectra cover the observed wavelength range of 
3800-8800 \AA. Total exposure times range from 1 to 6 hr per object. Results 
for the first two objects in this survey are presented by Cohen et al. (1995).

The object studied in most detail is 0226-1024 (Fig. 1), a bright 
BALQSO with $m_V=16.9$ and multiple troughs extending to at least 0.1c
blue-ward of the permitted emission lines. Its polarimetric features are 
representative of most objects in the sample, so it is the primary
focus of this paper. The continuum polarization (P) rises to 2\%
in the blue, and the polarized flux has a spectral index of -0.3. 
P dips across all of the broad emission lines, so they have considerably lower
polarization than the continuum.

P rises dramatically to $\sim$8\% in the C IV absorption trough and to 
$\sim$5\% in the Si IV absorption trough. The troughs are therefore shallower
in polarized flux than in total flux (with one exception, which is discussed 
below). This illustrates the importance of partial coverage of the continuum 
sources by the broad absorption line region (BALR), and the potential 
difficulties in determining the optical depth of the BAL troughs due to 
filling in by scattered light. A significant portion (8\%-100\%) of the total 
flux in the bottom of the C IV trough is scattered light.

\begin{figure}
\plotfiddle{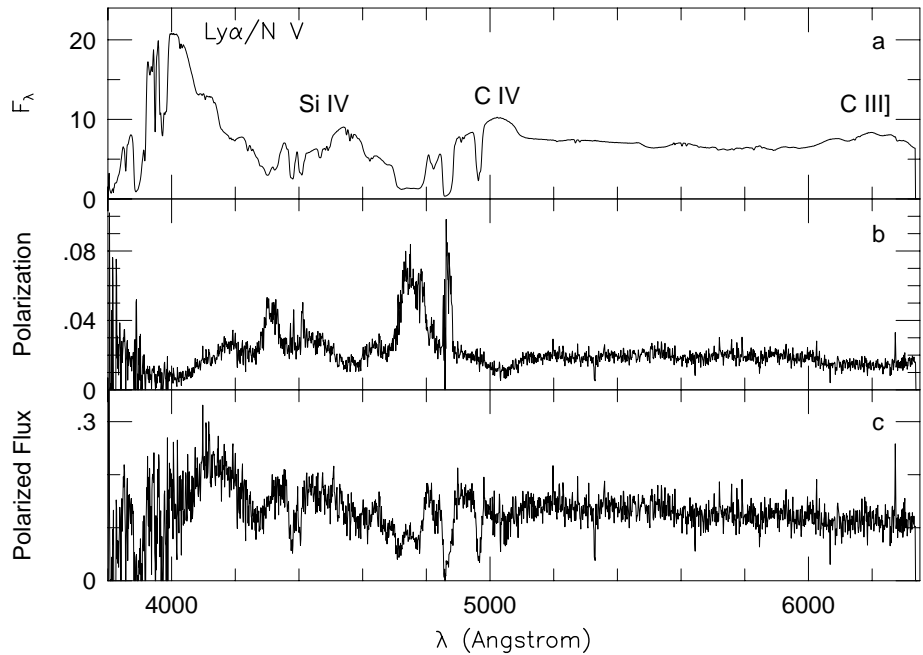}{3in}{0}{50}{50}{-200}{-20}
\caption{Keck spectropolarimetry of 0226-1024 (4 hr). (a) Total flux. (b)
 Fractional polarization. (c) Polarized flux (product of a and b). Total
 and polarized flux are multiplied by $10^{16}$.}
\end{figure}

\section{Relative Sizes}
There is much uncertainty in the absolute sizes of the various emitting 
and absorbing regions in BALQSO and QSO in general. There are some 
constraints from variability studies (Barlow et al. 1992; Kaspi et al. 1996), 
but they are not stringent. There are two constraints on relative sizes from 
spectroscopy. First, the continuum source must be smaller than the broad 
absorption line region (BALR) so that the BALR blocks the continuum emission 
(at least partially). Second, the N V BALR commonly blocks the Ly$\alpha$ 
emission line (e.g., Turnshek et al. 1988), indicating that the BALR is larger
then the broad emission line region (BELR).

Spectropolarimetry provides an additional constraint on the sizes of the
quasar emitting and absorbing regions. The low polarization of the broad
emission lines (BEL) relative to the continuum indicates that the BELR is 
extended, while the continuum source is compact relative to the electron 
scattering region. In other words, the electron scattering region is either 
smaller than or roughly cospatial with the BELR. The relative sizes of the 
various regions are summarized in Figure 2. Note that the BALR partially 
blocks the electron scattering region, leading to shallow absorption troughs 
in polarized flux. Continuum light is scattered, then absorbed, and not 
the reverse.

\begin{figure}
\plotfiddle{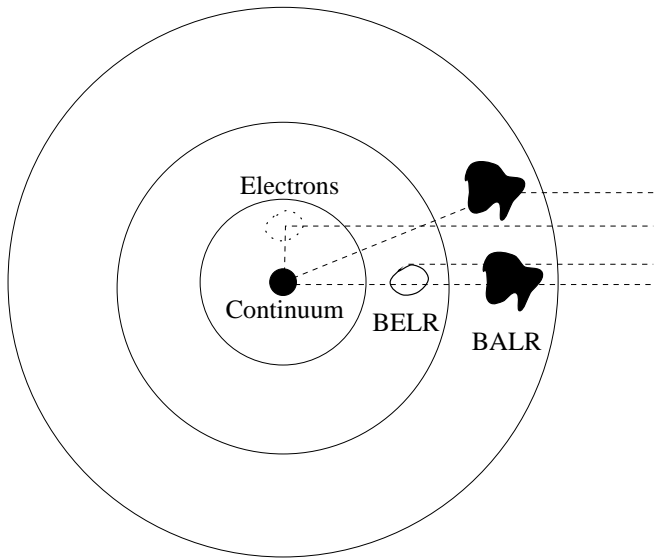}{3in}{0}{50}{50}{-110}{0}
\caption{Relative sizes of the nuclear emission/absorption regions.
         Direct and scattered rays are shown.}
\end{figure}

\section{Geometry}
There are a number of possible axisymmetric configurations of the scattering
region and BALR (Fig. 3). In the case where the BALR is an equatorial outflow 
(Figs. 3a and 3b), the QSO is viewed at high inclination to its symmetry axis.
The continuum shows relatively high polarization, by scattering from either a 
polar (Fig. 3a) or equatorial (Fig. 3b) distribution of electrons. If the 
scattering electrons are situated along the AGN axis, then any light 
resonantly scattered in the BALR is polarized perpendicular to the continuum 
light. On the other hand, if the scattering electrons are in the same plane as
the BALR outflow, then the resonantly scattered light is polarized at the same 
position angle as the continuum. In the first case, the polarization 
vectors cancel, giving a deficit of polarized flux, while in the second 
case, they add, giving a surplus of polarized flux. The argument is similar
if instead the BALR outflow is along the AGN axis (Figs. 3c and 3d). The main 
difference is that the degree of polarization is lower because the 
distribution of scatterers in the plane of the sky is more symmetric. Finally, 
it is possible that the BALR is arranged in a patchy spherical or asymmetric
geometry with low covering factor.

\begin{figure}
\plotfiddle{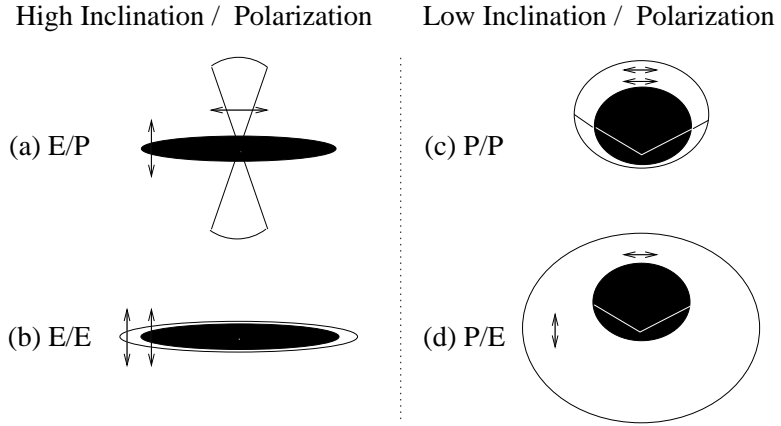}{2.5in}{0}{60}{60}{-140}{0}
\caption{BALR (filled) and electron scattering region (empty) geometries,
 viewed in the plane of the sky. Polarization vectors from the two regions 
 are shown. (a, b) Equatorial (E) BALR outflows viewed at high inclination 
 have high polarization. (c, d) Polar (P) BALR outflows viewed at low 
 inclination have low polarization.}
\end{figure}

Spectropolarimetry can help distinguish among the possible geometries 
of the BALR and electron scattering region. Discriminating between polar and 
equatorial outflow is problematic, since the main difference is in the 
magnitude of the continuum polarization. If the BALR outflow covers
about 12\% of the continuum source, then most non-BAL QSO are just BALQSO 
viewed on a line of sight not intersecting the BAL outflow (Weymann et al. 
1991). Few high-redshift non-BAL QSO have been observed with high polarimetric
accuracy, however they generally have lower polarization than BALQSO (see, 
e.g., Antonucci et al. 1996). This suggests that BALQSO are observed at 
greater inclination than non-BAL QSO, leading to the conclusion that the BALR 
outflow is equatorial.

Three of the objects in the sample show a deficit of polarized flux red-ward
of their permitted emission lines (0226-1024, CSO 755, and RS 23). Figures 
1 and 4 show this effect in N V, C IV and perhaps Si IV in 0226-1024. 
This could be due to either resonance scattering in the BALR or residual 
polarization of the BEL by electron scattering. If resonance scattering
is the cause and the BALR flow is equatorial, then the scattering electrons 
are distributed in a polar configuration. Hence, the geometry of Figure 3a is 
in best agreement with the polarimetric observations. Taking a closer look at 
the spectropolarimetry of 0226-1024 (Fig. 4), it is seen that in addition to
the deficit of polarized flux red-ward of C IV, there is a $+5\deg$ rotation 
in position angle (PA) which extends from $-14,000$ to $+14,000$ km s$^{-1}$, 
well outside the velocity range of the emission line. This rotation must be 
associated with resonantly scattered polarized flux from the BALR, and 
indicates a misalignment between the axes of the BALR outflow and the electron
scattering region. 

\begin{figure}
\plotfiddle{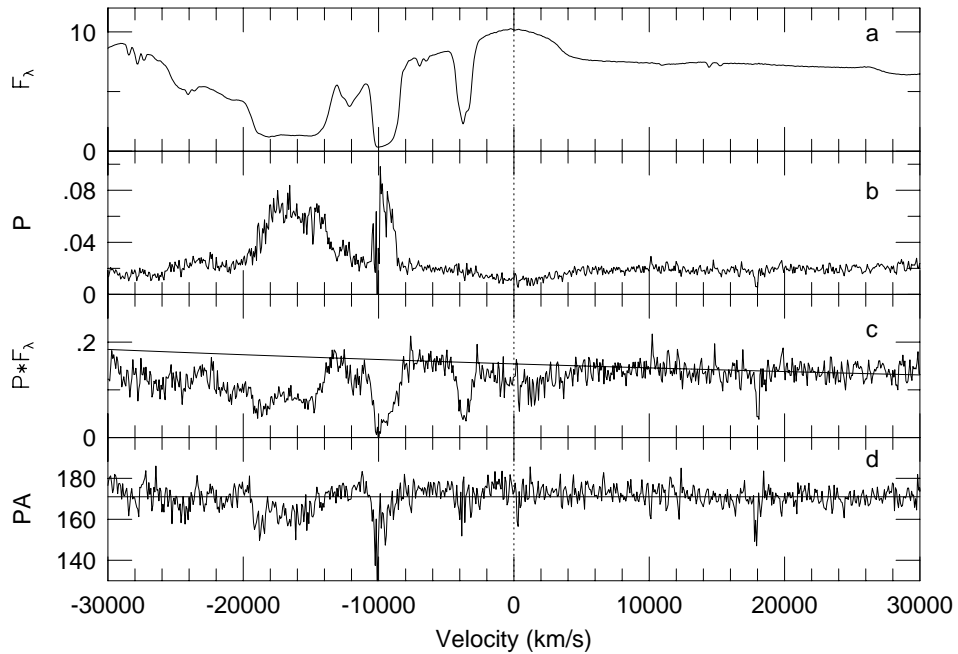}{3in}{0}{50}{50}{-200}{-20}
\caption{C IV trough of 0226-1024 (z=2.246). (a) Total flux. (b) Fractional 
 polarization. (c) Polarized flux with power law fit to the continuum. 
 (d) Position angle. The mean continuum PA is $171\deg$.}
\end{figure}

\section{Dynamics}

A common effect seen in BALQSO is greater polarization at low velocity than at
high velocity in each sub-trough (Fig. 4). This is easily explained if the 
electron scattering region is only partially covered by BALR clouds and 
coverage increases with outflow velocity (Fig. 5). If the BALR flow expands 
as it accelerates outward, then the higher velocity gas covers a larger 
portion of the electron scattering region, leading to lower polarization (for 
a given trough depth) at high velocities. 

The sub-troughs also show PA rotations of up to $-15\deg$ (Fig. 4), which are 
correlated to their velocity structure in polarized flux. This is naturally 
explained if the flow crosses the line of sight at an angle to the axis of the
scattering region (Fig. 5b). Note that the rotations in the troughs are 
opposite to the rotation due to resonance scattering in the BALR, allowing a 
distinction between the two effects. This is expected since BALR streamers 
crossing the line of sight absorb polarized light at the same PA that 
they scatter light from the central continuum source. The PA in the troughs is
driven away from the PA of the obscured electron scattered light. 

Another striking detail in the spectropolarimetry of 0226-1024 is the absence
of a polarization increase in the lowest velocity trough (Fig. 4). This 
suggests that it covers the electron scattering region almost completely. 
Perhaps it is due to BAL gas at a different radius or a completely different 
class of associated absorber. This is supported by the Keck HIRES data
of T. Barlow and  V. Junkkarinen (priv. comm.), which show that this trough 
breaks up into a number of sub-troughs at high resolution, whereas the other 
troughs remain smooth.

\begin{figure}
\plotfiddle{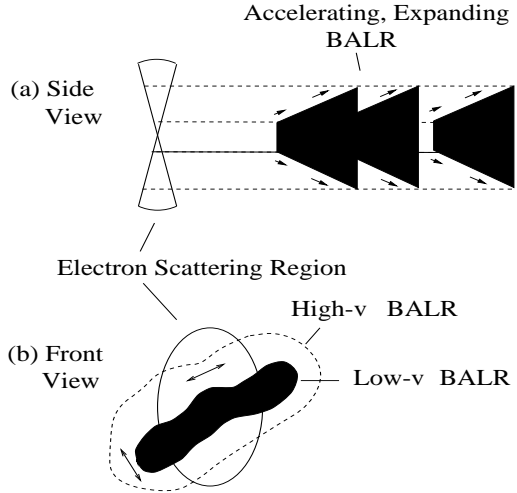}{2.5in}{0}{40}{34}{-140}{10}
\caption{(a,b) Higher velocity clouds in an accelerating, expanding outflow
   cover a larger portion of the electron scattering region . (b) A 
   misalignment between the BALR and electron scattering region axes causes 
   PA rotations.}
\end{figure}

\section{Conclusions}

BALQSO continuum light is polarized by electron (or dust) scattering from an 
extended region smaller than or the same size as the BELR, then is 
partially absorbed by clouds in the BALR. Resonantly scattered continuum light
from the BALR has been detected for the first time and is polarized 
roughly perpendicular to the continuum, suggesting that the electron 
scattering and BALR structures are perpendicular. If the BALR outflow is 
equatorial, then the electron scattering region lies near the polar axis. 
Velocity-dependent polarization and PA changes in the BAL troughs are 
consistent with an accelerating BALR outflow crossing the line of sight 
at an angle to the symmetry axis of the electron scattering region.

\end{document}